# HYDROGEOPHYSICAL CHARACTERIZATION OF TRANSPORT PROCESSES IN FRACTURED ROCK BY COMBINING PUSH-PULL AND SINGLE-HOLE GROUND PENETRATING RADAR EXPERIMENTS

Alexis Shakas[1], Niklas Linde[1], Ludovic Baron[1], Olivier Bochet[2], Olivier Bour[2] and Tanguy Le Borgne[2]

[1]Applied and Environmental Geophysics Group, Institute of Earth Sciences, University of Lausanne, Lausanne, Switzerland;

[2]Géosciences Rennes, Université de Rennes 1, Campus de Beaulieu, Rennes cedex, France;

Corresponding author: A. Shakas, Applied and Environmental Geophysics Group, Institute of Earth Sciences, University of Lausanne, 1015 Lausanne, Switzerland, alexis.shakas@unil.ch

**Key points:**

- Combined single-hole GPR and push-pull tracer tests help to infer transport phenomena
- Transport length scales are estimated through GPR difference imaging
- GPR inform on tracer velocity, fracture channeling and density driven flow

**Abstract**

The in situ characterization of transport processes in fractured media is particularly challenging due to the considerable spatial uncertainty on tracer pathways and dominant controlling processes, such as dispersion, channeling, trapping, matrix diffusion, ambient and density driven flows. We attempted to reduce this uncertainty by coupling push-pull tracer experiments with single-hole ground penetrating radar (GPR) time-lapse imaging. The experiments involved different injection fractures, chaser volumes and resting times, and were performed at the fractured rock research site of Ploemeur in France (H+ network, hplus.ore.fr/en). For the GPR acquisitions we used both fixed and moving antenna setups in a borehole that was isolated with a flexible liner. During the fixed-antenna experiment, time-varying GPR reflections allowed us to track the spatial and temporal dynamics of the tracer during the push-pull experiment. During the moving antenna experiments, we clearly imaged the dominant fractures in which tracer transport took place, fractures in which the tracer was trapped for longer time periods and the spatial extent of the tracer distribution (up to 8 meters) at different times. This demonstrated the existence of strongly channelized flow in the first few meters and radial flow at greater distances. By varying the resting time of a given experiment, we identified regions affected by density-driven and ambient flow. These experiments open up new perspectives for coupled hydrogeophysical inversion aimed at understanding transport phenomena in fractured rock formations.

# 1. Introduction

Characterization of flow and transport in fractured rock formations has been a central focus of hydrogeological research for several decades [*e.g., National Research Council,* 1996]. The interest arises from a multitude of applications, ranging from environmental remediation [*e.g., Andričević, and Cvetković*, 1996] to safe disposal of nuclear waste [*e.g., Cvetković et al.,* 2004]. Experimental studies over several scales is essential to constrain site-specific conceptual models [*Le Borgne et al.,* 2006] that can be used for numerical simulations of flow and transport for purposes of predictions, risk-assessment and decision-making [*Berkowitz,* 2002].

Fractured geological media are often assumed to be scale-invariant and power-law distributions have been successfully used to statistically describe scaling properties in terms of length or aperture [*Bonnet et al.,* 2001]. The resulting flow heterogeneity is characterized by strong channeling at both fracture and network scales, which implies that advection occurs mainly through preferential paths. This directly affects transport processes that display strongly non-Fickian dispersion behavior, including early tracer breakthrough and late time tailing, as observed in tracer test experiments [*Haggerty,* 2000; *Becker and Shapiro,* 2003; *Berkowitz,* 2002]. The physical interpretation of these observations is often ambiguous as different physical processes, for example, diffusive trapping into a low velocity zone or advection into variable velocity channels, may cause similar tailing behavior in tracer breakthrough curves [*Kang et al.,* 2015]. This motivates the development of geophysical approaches that enable observing the motion and spatial distribution of tracers in situ. This is the focus of the present study.

The complexity of tracer pathways in fractured systems implies that tracer recovery, in classical cross-borehole tests, decays fast as the distance between injection and observation wells is increased. On the other hand, push-pull experiments [*Istok,* 2012]—also called single-well injection withdrawal experiments—with their higher tracer retrieval offer a time and cost-effective method for obtaining information about transport properties of fractured rock formations [*Nordqvist and Gustafsson,* 2002]. Push-pull experiments are carried out by packing-off a fractured section of the borehole using a dual-packer system and injecting (pushing) a tracer (or multiple tracers) into the fractured system through the packed-off interval (injection interval). After pushing the tracer into the system, a chaser fluid is used to clean the injection chamber and to push the tracer further out into the formation. The chaser fluid is usually water at ambient aquifer conditions and the volume of chaser injected varies depending on the scale of investigation. During the pulling period the flow is reversed to produce a tracer breakthrough curve.

The expected reversibility of advective transport in push-pull experiments implies that the recorded tracer breakthrough curve is only weakly sensitive to the transmissivity distribution of the fractured rock formation, and that the mean arrival time of the tracer breakthrough curve is uninformative [*Becker and Shapiro*, 2003]. Indeed, push-pull experiments are mainly sensitive to irreversible time-dependent processes [*Nordqvist and Gustafsson,* 2002*; Kang et al.,* 2015] such as sorption and diffusion. It is possible to increase the sensitivity of push-pull experiments to these processes by introducing a resting time between the pushing and pulling phases [*Berkowitz,* 2002]. Multiple push-pull tests with varying pushing and/or chasing volumes make it possible to engage different volumes of the system and gain scale-dependent information [*Gouze et al.,* 2008]. The results can be compared to analytical or numerical solutions, thus allowing inference about flow and transport properties of the system [*Becker and Shapiro,* 2003; *Le Borgne and Gouze,* 2008].

Geophysics offers high resolution and high spatial coverage data that complement the information obtained by hydrological experiments [e.g., *Hubbard and Linde*, 2011]. For instance, ground penetrating radar (GPR) reflection imaging makes it possible to image dynamic processes and associated length scales of hydrogeological experiments in fractured rock formations, particularly when tracers with sufficient contrast in electrical conductivity are used, such as saline tracers [*Tsoflias and Becker,* 2008]. The unique ability of GPR reflection imaging to remotely detect fractures of millimeter-thin aperture has been rigorously investigated in both laboratory [*Grégoire and Hollender,* 2004] and field-based conditions [e.g., *Dorn et al.,* 2012b]. Recently, surface [*Becker and Tsoflias,* 2010], cross-well [*Day-Lewis et al.,* 2003] and single-well [*Dorn et al.,* 2012a] GPR have been used in conjunction with dipole saline tracer experiments, not only to dynamically image the migration of tracer movement between boreholes but also to condition discrete fracture network (DFN) models [*Dorn et al,* 2013].

DFN models are conceptual models used to describe fractured rock formations, where flow and transport only occurs inside fractures [e.g.*, de Dreuzy et al.,* 2012]. This approximation is particularly suitable for crystalline rock formations [*e.g., Nordqvist and Gustafsson,* 2002]. To adequately condition DFN models it is necessary to describe the physical properties of the fractures at both the single fracture and the fracture network scales [*de Dreuzy et al.,* 2012]. For a single fracture this description can include the mean aperture, fracture orientation and fracture length. It is practically impossible to gain independent information on these properties through push-pull tests alone since the length scale over which the tracer is transported is unknown and fracture orientation is only known at the borehole location. Moreover, the effect of ambient flow and the buoyancy effect of tracers with a significant density contrast with respect to the formation water, are usually ignored when interpreting push-pull data.

To address the inherent ambiguity in interpreting push-pull tests alone, we conducted a series of combined push-pull and single-hole GPR experiments in a crystalline aquifer. We performed repeated push-pull experiments with varying resting times and chasing volumes and monitored the GPR response with both fixed and moving antenna configurations. In this contribution we investigate how GPR monitoring of push-pull experiments can provide constrains on flow and transport characteristics of the fractured system.

**2. Field Site and Experimental Setup**

The experiments were carried out between the 19$^{th}$ and 26$^{th}$ of June 2014 at the Ploemeur fracture rock experimental site in Brittany, France (H+ network of experimental sites). Previously acquired data at this well-studied site, such as optical, acoustic and geological logs can be found in the H+ observatory database (http://hplus.ore.fr/en). The aquifer supplies drinking water to the town of Ploemeur (20,000 inhabitants) and it is mainly composed of granite and mica schists [*Ruelleu et al.*, 2010]. At the borehole scale, only a few fractures dominate the hydraulic behavior of this highly transmissive aquifer (the average transmissivity at site scale is around $T$=10$^{-3}$ m$^2$/s [*Le Borgne et al.*, 2004, 2006]). Borehole GPR data have been previously acquired at the experimental site named Stang-Er-Brune that is located 3 km west of the water supply wells [*Le Borgne et al.*, 2007; *Dorn et al.*, 2011, 2012a, 2012b; *Kang et al.*, 2015]. To the best of our knowledge, this is the first time that push-pull tests and single-hole GPR are combined in a field experiment.

A series of cross-borehole and push pull tracer tests were previously performed with fluorescent dyes on this site [*Kang et al.*, 2015]. Breakthrough curve tailing was found to be characterized by power law behaviors, $c(t) \sim t^{-1-b}$ at late times, with $b$ ranging from 0.75 to 1. The tailing exponents were systematically larger (which implies that tailing was less significant) under push pull conditions than under cross-borehole conditions. This implies that matrix diffusion is not expected to be significant in this low permeability granite and that the dispersion behavior measured from the fluorescent dye experiments was mostly controlled by dispersion in fracture planes, as observed in the similar geological setup at the Mirror Lake site [*Becker and Shapiro*, 2003].

The injection (B1) and monitoring (B2) boreholes are separated on average by 6 m. We conducted push-pull experiments in B1 while we performed single-hole GPR monitoring in B2. In contrast to previous single-hole GPR acquisitions [e.g., *Dorn et al*, 2012a] we used a flexible liner (blank flexible liner by FLUTe) to seal-off the GPR monitoring borehole from the aquifer. The liner was filled with formation water, extended to the bottom of the borehole and allowed smooth movement of the GPR antennas while preventing conductive tracer from entering the borehole and thereby affecting the effective GPR antenna signal (see *Dorn et al.* [2011, 2012a]). This also minimized the propagation of pressure variations arising from the antenna movement to the surroundings. A schematic description of the experimental setup is shown in Figure 1.

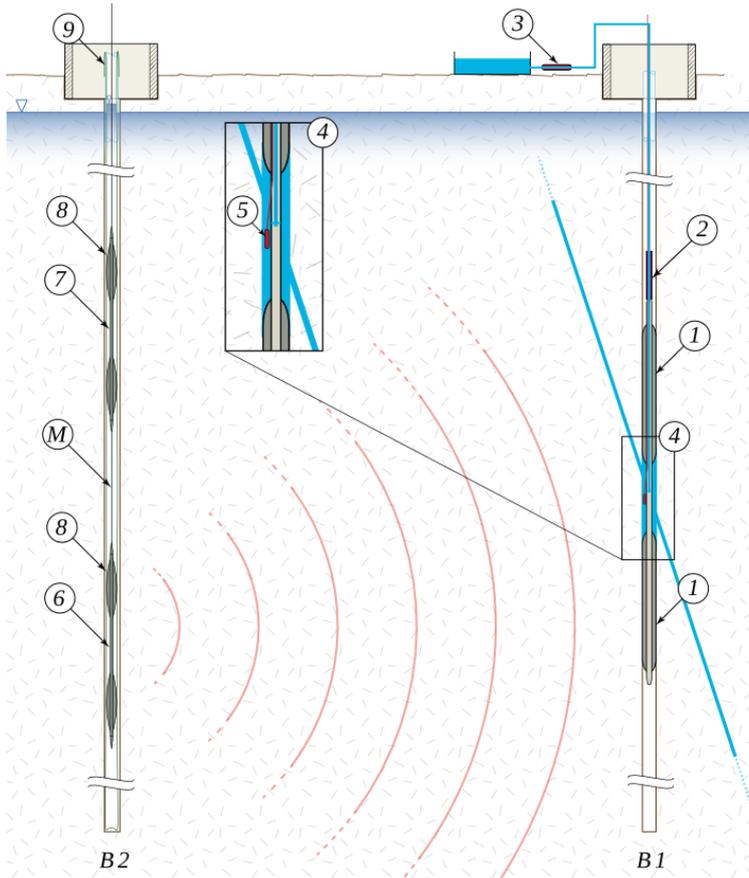

**Figure 1:** Schematic description of the combined push-pull experiment in borehole B1 and single-hole GPR monitoring in borehole B2 (not to scale). In B1, a section of the borehole is isolated using two inflatable packers (1) while a pump (2) and a flowmeter (3) are used to control the pushing and pulling of saline tracer in the injection interval (4). A CTD diver (5) measures the electrical conductivity, temperature and hydraulic pressure in the chamber. In B2, GPR monitoring takes place with the transmitter (6) and receiver (7) antennas, which have a fixed separation with midpoint M and are centered in the borehole using custom-made centralizers (8). B2 is hydraulically isolated from the surroundings by a flexible borehole liner (9).

We present five push-pull tests with injections at either 50.9 m or 78.7 m (see Table 1). We focus the analysis on experiments I, II and III; experiments IV and V are only used herein assess the repeatability of our results. Previous studies by *Le Borgne et al.* [2007] and *Dorn et al.* [2012a, 2013] concluded that the fracture at 50.9 m in B1 connects to low-permeable, sub-horizontal fractures while the fracture at 78.7 m connects to transmissive, sub-vertically oriented fractures.

These fractures were interpreted to be of the order of ~10 m in length and to have apertures in the millimeter to sub-millimeter range. We used a custom-made double-packer in B1 to seal off the injection interval during the push-pull experiments (Figure 1). The double-packer consisted of two inflatable packers separated by 0.6 m, which were inflated to firmly adhere on the borehole wall. We monitored and manually controlled the pumping rate with a flowmeter and pump (MP1 by Grundfos) installed above the double packer, while we measured the electrical conductivity, temperature and pressure in the injection interval using a CTD diver by Schlumberger. Halfway through each saline tracer injection we injected an additional fluorescent conservative tracer that had the same salinity as the formation water. In the present study we focus on the joint analysis of GPR and saline tracer breakthrough curve data.

**Table 1:** Push-pull and GPR parameters for experiments I to V.

| Experiment | I | II | III | IV | V |
|---|---|---|---|---|---|
| Injection Depth (m) | 78.7 | 78.7 | 50.9 | 78.7 | 78.7 |
| Monitoring depths (m) | 75.1 | 50-90 | 30-60 | 50-90 | 50-90 |
| GPR Antennas | Fixed | Moving | Moving | Moving | Moving |
| Tracer density (kg m$^{-3}$) | 1044 | 1044 | 1044 | 1042 | 1041 |
| Salinity (g kg$^{-1}$) | 44 | 44 | 44 | 42 | 41 |
| Tracer Volume (L) | 100 | 100 | 100 | 100 | 100 |
| Chaser Volume (L) | 90 | 90 | 100 | 90 | 710 |
| Resting Time (hh:mm) | 00:00 | 00:00 | 04:45 | 03:47 | 00:00 |
| Pumping rate (L min$^{-1}$) | 3.2 | 3.2 | 3.3 | 3.3 | 3.3 |
| Mass Recovery (%) | 55 | 51 | 79 | 40 | 27 |

For the GPR acquisitions we used the RAMAC 100 MHz system with slimhole borehole antennas, manufactured by MALÅ. In all experiments we used a fixed separation of 4 m between the midpoints of the transmitter and receiver antennas. We centered the antennas in the borehole by using custom-made flexible centralizers that we attached to the top and bottom of each antenna (Figure 1). The use of centralizers and a flexible liner made the GPR acquisition very smooth and the estimated positioning errors during the moving antenna experiments were minor (standard deviation of 8 mm). During the moving antenna experiments we acquired data with a spatial sampling rate of 5 cm along the borehole. For the moving antenna experiment, we sampled at 1148 MHz for a time-window of 450 ns, and used 64 stacks per trace to increase the signal-to-noise ratio. For the fixed antenna experiment, we used a sampling rate of 4278 MHz and 128 stacks per trace.

While GPR reflections contain information about the fracture aperture at the mm scale [*Dorn et al.*, 2011], the spatial resolution along the fracture plane (based on the Fresnel zone) is roughly 1 m for the 100 MHz antennas we used and the distances we investigate [e.g., *Jol,* 2009, sec. 1.3.4]. After the processing of the moving antenna experiments, the resolution is refined to roughly a quarter of the wavelength (i.e., 0.25 m). Both types of acquisition (fixed and moving antenna) resulted in data with very high signal-to-noise ratios that required minimal processing.

**3. GPR Data Processing Steps**

In this section, we describe the processing steps that were applied to the raw single-hole GPR data. We begin with the basic concept of single-hole GPR and continue to analyze each processing step (see *Jol* [2009] for a more detailed description of GPR theory and processing).

*3.1. Basic concepts of borehole GPR*

A dipole GPR transmitter antenna, as the one we used, generates an alternating high-frequency electric field. This field propagates symmetrically around the antenna axis but is the strongest in the plane that is perpendicular to the antenna and intersects the antenna midpoint [*Jol,* 2009, sec. 1.4].

The propagating field is attenuated and dispersed as it travels through the rock matrix and water-filled fractures. Reflections occur at rock-fracture interfaces due to the high contrast in electrical properties between the rock matrix and water. The strength of the reflections depends on fracture properties such as orientation, aperture and roughness. Within the water-filled fracture the electric field experiences strong attenuation and dispersion, which results in a decreased amplitude and a phase shift of the electric field. When the electrical properties of the fracture filling change, such as the change in electrical conductivity induced by the presence of a saline tracer, the magnitude and phase of the reflected electric field also change [*Bradford and Deeds,* 2006*; Tsoflias and Becker,* 2008].

The receiver antenna measures a voltage as a function of time (a GPR trace) that is proportional to the amplitude of the reflected electric field. If the GPR antennas are kept fixed at a specific depth along the borehole, changes in the reflected traces over time are indicative of salinity variations in the vicinity of the transmitter-receiver pair. These changes reflect the (averaged) temporal dynamics of the saline tracer in the region surrounding the transmitter-receiver midpoint.

If the GPR antenna pair is moved along the borehole during a push-pull experiment—using saline or any other tracer with electrically contrasting properties to the ambient fluid—then the changes in the GPR traces recorded along the borehole can be directly linked to the spatio-temporal migration of the tracer.

*3.2. Time corrections, filtering and normalization*

The initiation time (time-zero) of the receiver and transmitter antennas varies slightly every time the system is turned on and it may drift over time. To account for this, we applied a time-zero correction following *Peterson* [2001]. After the time-zero correction we applied a digital bandpass filter (a Kaiser Window with edges at 10-40-150-200 MHz) to concentrate on frequencies around the dominant frequency that is indicated by the antenna manufacturer as 100 MHz.

In single-hole GPR acquisition, part of the electric field is refracted at the GPR borehole wall [*e.g., Dorn et al.,* 2011]. These refractions (termed borehole refractions in the following) of high amplitude are measured earlier in time than the reflections from the tracer-filled fractures that are located further away. By sealing the GPR borehole with the flexible liner we ensured that the borehole conditions remained constant during acquisition, which implies that the amplitude and phase of the borehole refractions were constant over time at each depth along the borehole.

After the initial time-zero correction, subtle time shifts (with a mean of 0.4 ns and a standard deviation of 0.25 ns) were observed in the borehole refractions when repeating the measurements at the same locations in the borehole. We attribute these time shifts to small variations in the initiation time of each pulse and to cm-scale vertical and/or horizontal shifts in the location of the antennas within the borehole. To correct for these time shifts we compared collocated traces to a reference trace that was measured at each depth before the initiation of the push-pull experiment. Since these time shifts were smaller than the sampling rate, we upsampled each trace at 5 times the original sampling rate by performing a Fourier transform on the data and padding the Fourier spectrum with zeros. We then shifted the upsampled traces so that the borehole refractions coincided with the ones of the reference trace. We subsequently performed an inverse Fourier transform and downsampled the data to the original sampling rate to obtain the resulting phase-shifted traces.

*3.3. Time-lapse changes and migration*

Before each push-pull experiment, the measured GPR data were indicative of the ambient conditions (e.g., water-filled fractures with formation water, the presence of boreholes and double-packer system). After the initiation of the push-pull experiment, the saline tracer started filling the fractures that are connected to the injection interval. The electromagnetic field that is reflected by the tracer-filled fractures becomes amplitude and phase shifted. These changes are subtle and are best appreciated by subtracting the later traces from the reference trace. To minimize the errors in the subtraction process, it is important to precisely locate the traces along the borehole and to align the traces with the reference (see section 3.2).

Figure 2 displays the reference trace and a trace recorded during a push-pull experiment. We also show the difference of the later trace with respect to the reference trace (difference trace). The difference trace has clear peaks that accentuate the changes that result from the presence of the saline tracer.

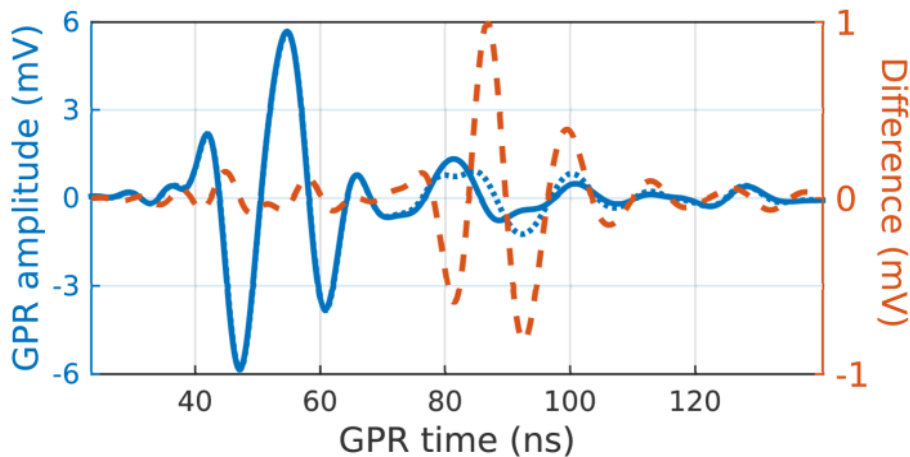

**Figure 2:** GPR data from experiment I. The reference trace (solid blue line) and a trace measured after pushing (dotted blue line) are shown together with the difference of the two traces (dashed red line). The difference is plotted on a different scale to the right of the figure, that is 6 factors smaller than the initial traces.

To partly account for attenuation of the signal [*Jol,* 2009, sec. 1.6.2], we applied a gain to scale the later times in the difference traces with respect to the earlier times. We choose to linearly scale each GPR difference trace with time, with appropriate scaling so that the resulting amplitudes are normalized to 1.

As a final step, we performed migration that refers to the summation of energy along certain paths to account for spherical spreading. A collection of traces along the borehole forms a section. We refer to sections from which the reference is subtracted as difference sections. We migrate the difference sections using a Kirchoff-migration algorithm developed for seismic data from the CREWES Matlab toolbox (http://www.crewes.org/) under the assumption of a constant GPR velocity of 0.12 m ns$^{-1}$. *Dorn et al.* [2012a] showed through extensive cross-hole travel time tomography that the radar velocity in the granite is practically constant and isotropic.

Figure 3a illustrates a synthetic 3D-model with the electric field at a given instant in time and the resulting saline tracer distribution (at the end of the injection period) within a rectangular, tilted fracture of uniform (0.5 mm) aperture. For this model, we calculate the GPR forward response using a newly developed algorithm [*Shakas and Linde*, 2015] to simulate the section prior to (Figure 3b) and at the end of the tracer injection (Figure 3c). The finite emission time (some 20 ns in this example) of the alternating source manifests itself as positive and negative signal amplitudes in the GPR traces. This effect is visible in the difference section (Figure 3d) and the migrated difference section (Figure 3e). The finite emission time also implies that the actual fracture location coincides with the shortest distance at which the absolute amplitude of the migrated difference section is above zero (see Figure 3e). Indeed, the fracture has an aperture of 0.5 mm, but the corresponding reflection in the migrated difference section shows an imprint of the source over a width of approximately 2 m.

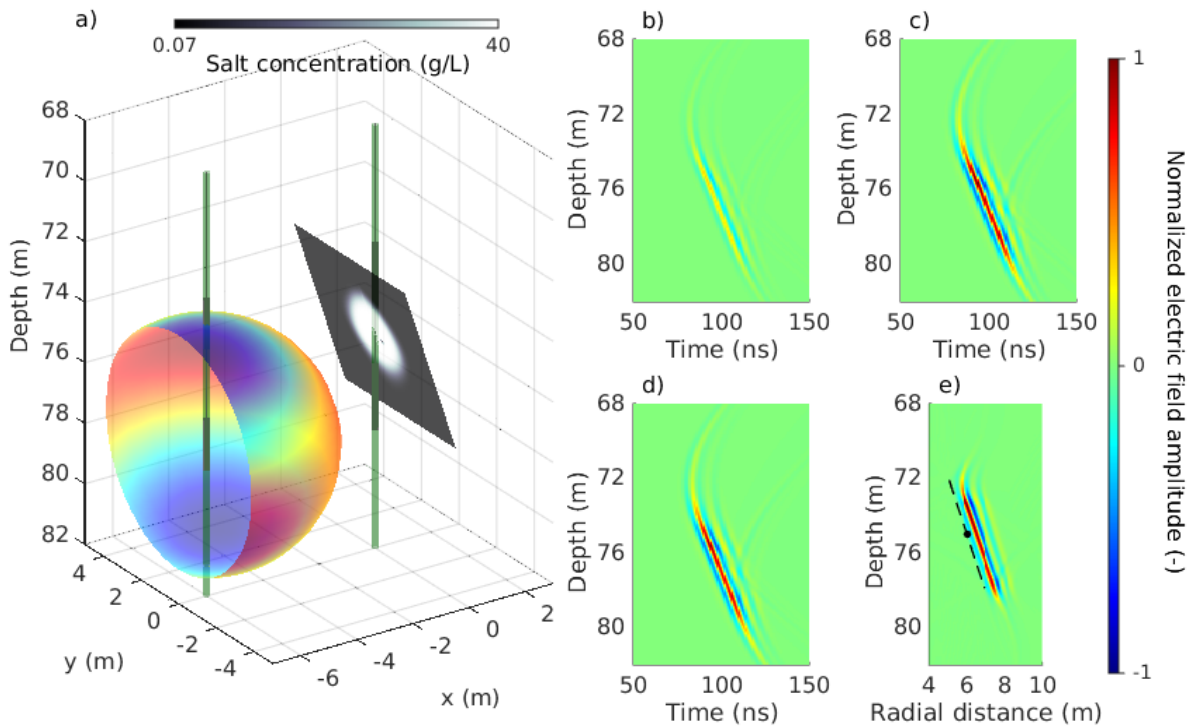

**Figure 3:** (a) A synthetic 3D model with the electric field and the tracer concentration in a fracture at a given time. (b) The simulated GPR section with formation water, (c) the same section with the tracer distribution (shown in (a)), (d) their difference and the migrated difference section, along with the fracture location and injection point (e). The GPR simulations were performed with a newly developed algorithm [*Shakas and Linde*, 2015] and the flow-and-transport simulation was done with MaFloT 2D.

## 4. Results

### 4.1. Fixed antenna acquisition

For the fixed antenna acquisition (experiment I in Table 1) we kept the midpoint between the transmitter and receiver antennas at 75.1 m depth while the injection depth was 78.7 m. We chose this location based on results from a previously acquired moving antenna experiment that suggested an upward tracer migration.

First, we computed the Root Mean Square (RMS) of each GPR difference trace and plotted the evolution of the RMS curve over time along with the measured electrical conductivity in the injection interval (Figure 4). This measure is representative of the change in the total (measured) reflected energy and is, therefore, indicative of the volume and electrical conductivity of the saline tracer within the antenna range. In the following, we refer to this type of curve as the GPR breakthrough curve.

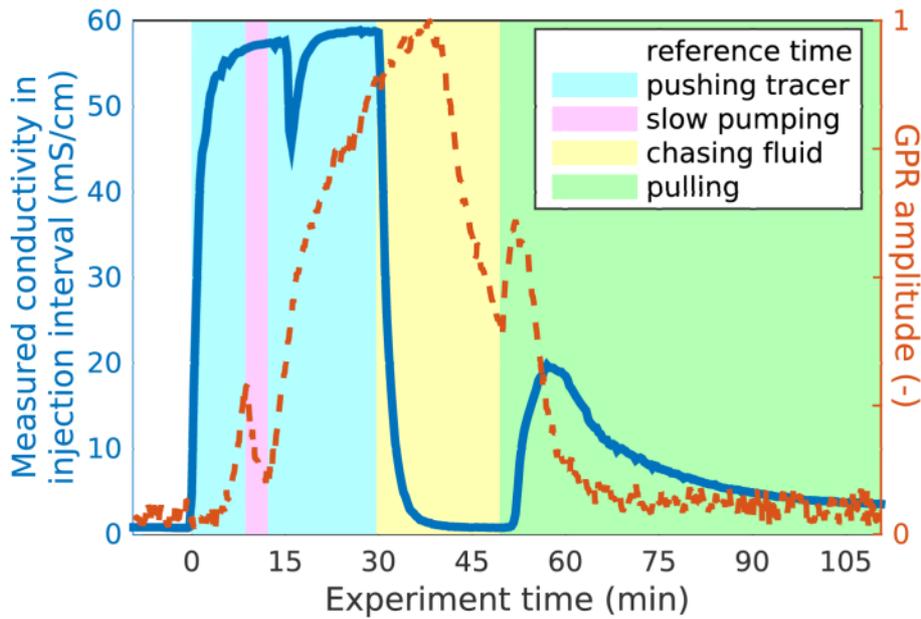

**Figure 4:** GPR and push-pull data for experiment I (fixed-antenna). The measured conductivity in the injection interval is shown (solid blue line) along with the normalized GPR breakthrough curve (dashed red line).

Before the push-pull experiment started (reference acquisition) the GPR breakthrough was close to zero and dominated by noise. The saline tracer reached the injection interval at $t = 0$ min and the magnitude of the GPR breakthrough increased considerably about 5 min later. One reason for this delay is that the antenna system was located above the injection depth and the tracer needed to travel some distance upwards before it was detected. Also, a sufficient volume of tracer must be injected to allow measurable changes in the reflected GPR signal.

At $t$ = 10 min after the saline tracer reached the injection interval, we temporarily paused the saline tracer injection and started to inject a second tracer of ambient conductivity during 3 minutes at 0.33 L min$^{-1}$, that is, at a 10 times lower rate than during the saline tracer injection. This period is indicated in Figure 4 as 'slow pumping'. Note that the second tracer was injected at the surface and it took about 3 min to reach the injection interval (seen as a decrease in the measured conductivity at $t$ = 15 min). The GPR breakthrough curve responded to the lower injection rate by a sharp decrease of its amplitude that only starts to increase after the previous flow rate and saline tracer injection was resumed.

The peak in the tracer breakthrough curve was reached at $t$ = 25 min and the GPR breakthrough curve shows a clear maximum at $t$ = 38 min. After the peak was reached in the GPR breakthrough curve, the amplitudes start decreasing as the tracer was diluted by the addition of the chaser fluid and migrated away from the antenna location.

During the pulling phase, another peak was observed in the GPR breakthrough curve at $t$ = 52 min that is smaller in amplitude and duration. The peak in the pulling phase was reached later in the tracer breakthrough than in the GPR breakthrough because the tracer first passed by the static antenna setup before migrating further down towards the injection interval, at 78.7 m, before it reached the CTD diver in the injection interval. At times later than $t$ = 70 min, the GPR breakthrough curve showed noisy behavior at an amplitude that was more than twice the one measured before the tracer injection.

*4.2. Moving antenna acquisition*

During the moving antenna acquisitions – see experiments II to V in Table 1 – we monitored over a depth range that was wide enough to capture all expected changes induced by the saline tracer. Here we present experiment II that had the same experimental parameters as experiment I, apart from the moving antenna setup. The difference sections after migration are shown for this experiment, along with the acquisition times, in Figure 5. Note that we only present the part of the depth section where temporal changes are visible. The migration of the saline tracer from the injection location through an upward trajectory of roughly 8 m is clearly seen. The maximum extent of the tracer is found in Figure 5h, which corresponds to a difference section acquired at the end of the chaser injection. The sharp change in the distribution of tracer between pushing and pulling is most evident by comparing the difference sections obtained right before (Figure 5h) and right after (Figure 5i) the initiation of the pulling phase. There are little indications of significant amounts of remaining tracer at the end of the experiment (Figure 5j). Throughout the experiment, a fraction of the tracer remained close to the injection location.

In Figure 6 we plot the tracer breakthrough curve along with the GPR breakthrough curves at three depths that are indicated in Figure 5j. The three depths were chosen such that one is close to the injection depth, one is the same as the depth during the fixed antenna acquisition (see section 4.1) and one is close to the furthest extent of the tracer imaged by the GPR. In the moving antenna configuration, GPR measurements at a given depth were repeated every ~5 min. As expected, the noise levels in the GPR breakthrough curves were noticeably higher than for the fixed antenna acquisition—around 0.3 normalized units for the moving acquisition compared to 0.05 normalized units for the fixed acquisition—but a clear signal related to tracer transport was still evident

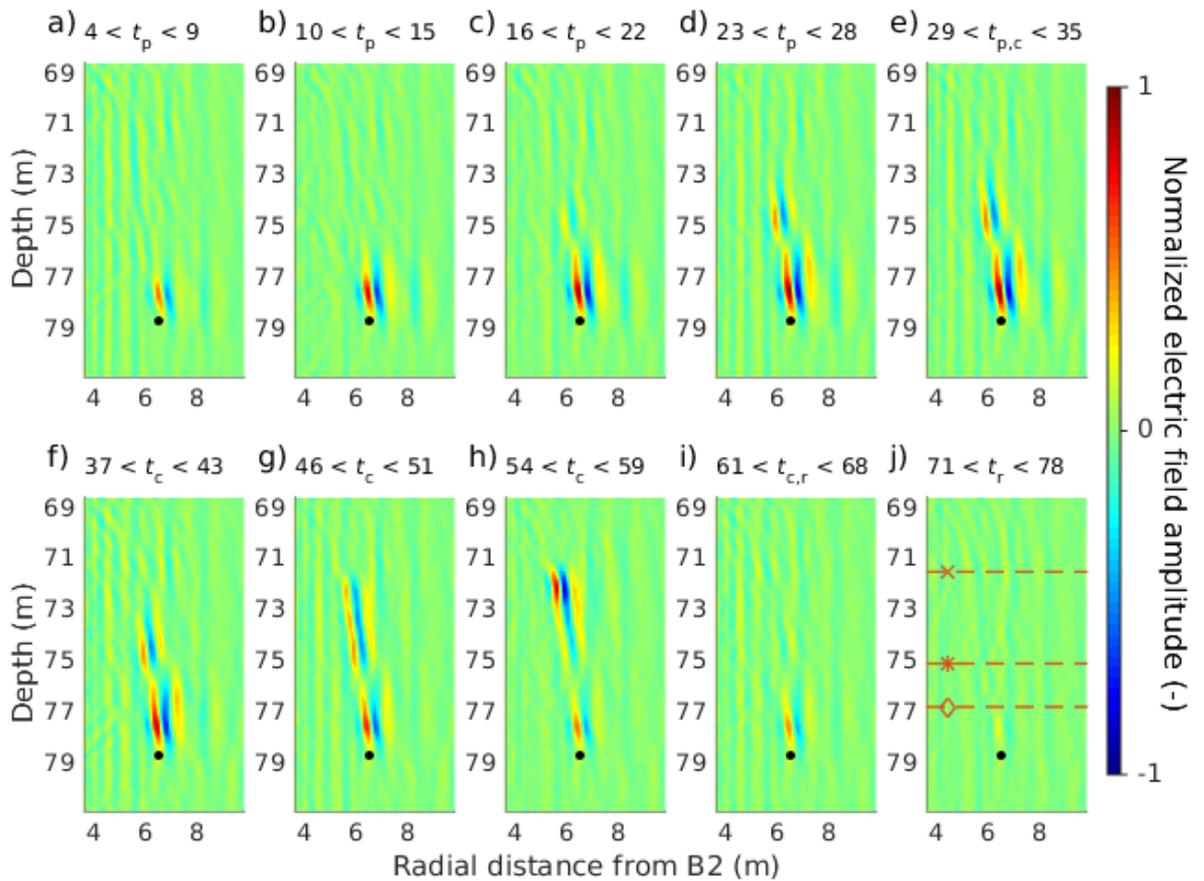

**Figure 5:** Migrated difference sections from experiment II. The black dot indicates the injection location and the acquisition times for each section (in min) are shown at the top of the sub-figures along with the phase of the push-pull experiment indicated by the index (*p*:pushing, *c*:chasing, *r*:pulling). The dashed red lines in 5j are analyzed in Figure 6.

Near the injection location at 76.8 m depth we measured strong difference signals towards the end of the saline tracer injection that decreased during the chasing period but prevailed until the end of the pulling period, suggesting that saline tracer was still present at this depth. The GPR breakthrough curve was considerably above the noise level from $t = 5$ min to $t = 68$ min. At the depth of 75.1 m (same as the fixed antenna experiment) we measured a peak with smaller amplitude that extended from $t = 15$ min to $t = 55$ min. Finally, the GPR breakthrough curve measured at 71.5 m reached a well-defined peak at $t = 56$ min. The saline tracer did not remain for a long time at this depth, but a considerable volume must have reached this depth interval to account for the strong changes in reflectivity (also seen in the surrounding traces in Figure 5h).

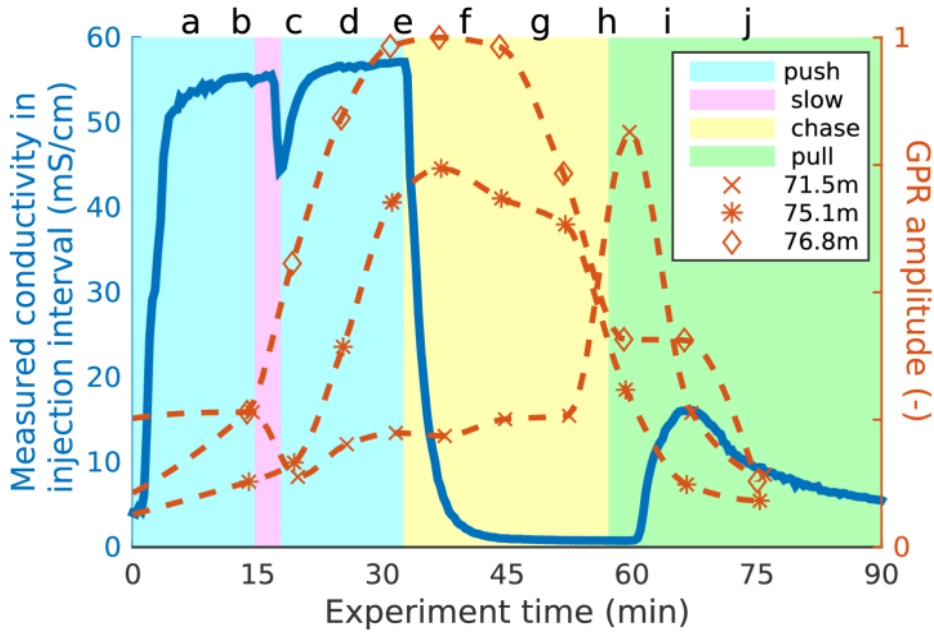

**Figure 6:** Plot of the measured conductivity in the injection interval (solid blue line) along with the GPR breakthrough curves (dashed red lines) during experiment II. The GPR breakthrough curves are shown for three depths (see Figure 5j). The letters at the top of the figure refer to the times at which the GPR profiles were taken (see the difference sections in Figure 5).

*4.3. Imaging ambient transport processes during the resting time*

We now present the results of experiment III that was performed with the saline injection at 50.9 m (see Table 1), a moving antenna acquisition and a resting time (4:35h) between pushing and pulling. This experiment highlights density related effects. As before, we only show the depth interval where strong temporal changes occurred. We have no GPR data between $t = 180$ min and $t = 270$ min as we charged the antenna batteries during this time period.

We present the difference sections from this experiment in Figure 7 and focus on two regions: an *upper* region found above the injection depth and a *lower* region found below the injection depth (shown in Figure 7a). The migrated difference sections show that changes during the pushing period arise mainly in the upper region and in the vicinity of the injection location (Figs. 7b and 7c). During the chasing period, from $t = 30$ min to $t = 50$ min, some of the tracer migrated to the upper region from 51 m to 47 m depth and some to the lower region from 52 m to 54 m depth (seen in Figs. 7d and 7e). At the beginning of the resting period, some tracer remained in the vicinity of the injection location (3 m radial distance and 51 m depth in Figure 7f). During the pulling phase (Figure 7i), the tracer from the upper region is recovered very quickly while the tracer in the lower region takes more time to be recovered. The tracer breakthrough curve is shown in Figure 7k.

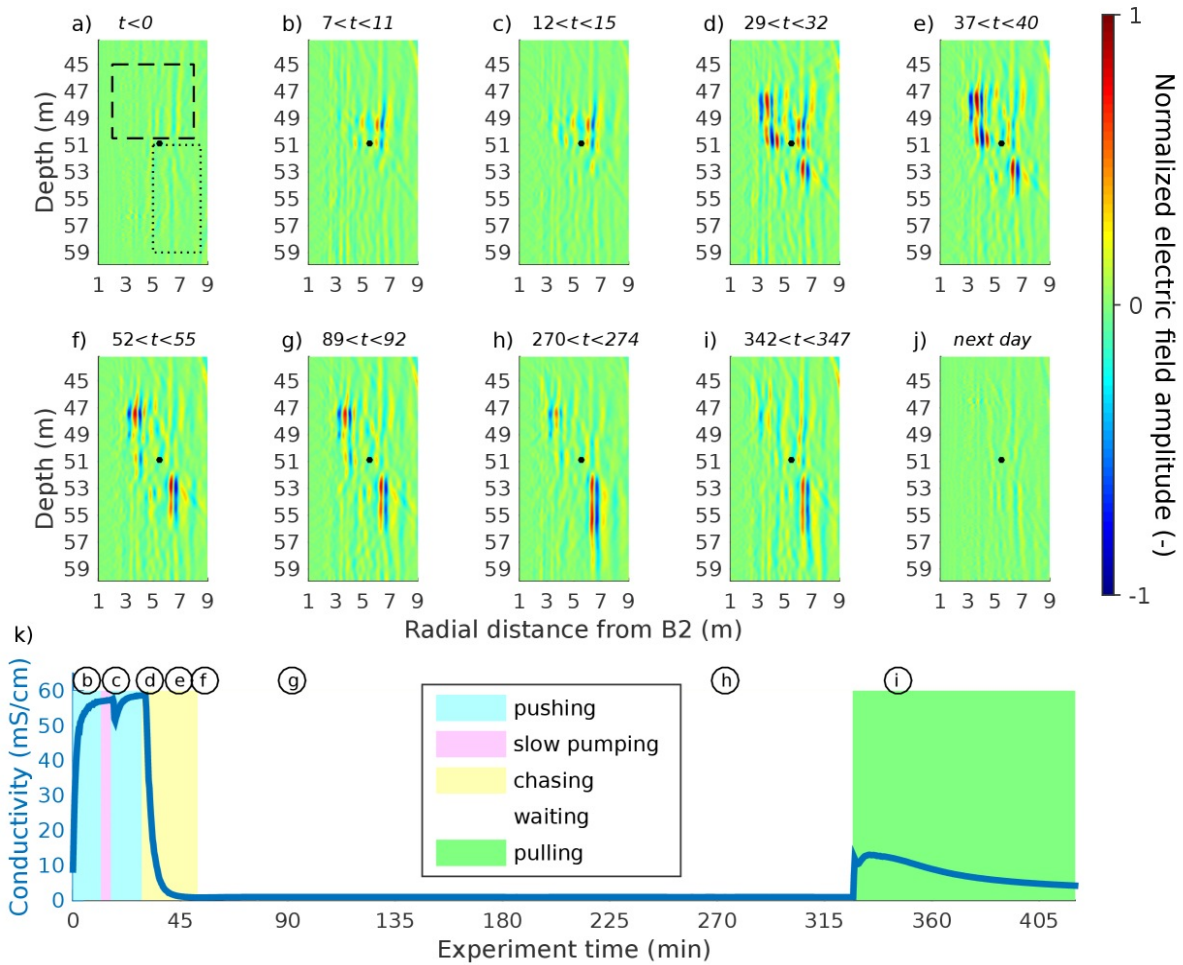

**Figure 7:** Figures (a) to (j) show the migrated difference sections from experiment III. The black dot indicates the injection location and the acquisition times for each section (in min) are shown at the top of the sub-figures and also indicated along with the measured conductivity in the injection interval (k).

The two regions of interest showed considerably different responses. The upper region shows amplitude differences at early times that were strongly attenuated during the resting time, and almost disappeared once the pulling begun. The lower region showed considerable amplitude differences that arise at the end of the pushing period (Figs. 7d to 7i). The saline tracer that caused this change exhibits a strong downward migration during the resting time, from 55 m (Figure 7f) to 59 m (Figure 7h).

In Figure 8 we plot the GPR breakthrough curves for the complete depth range at which changes were visible during the experiment. The injection depth and the two areas of interest (upper and lower region) are indicated. As was evident in the migrated sections of Figure 7, we can see the upward and downward migration of the tracer from the injection depth. In this figure, the downward migration of the tracer during the resting time is evident from $t = 60$ min and $t = 330$ min. During the resting time and after we charged the batteries, we measured weaker GPR differences in the top than in the lower region (see Figure 8 between $t = 270$ min and $t = 330$ min). During the pulling phase, the GPR breakthrough curves in the upper region were quickly attenuated while the ones in the lower region showed considerably high values throughout the acquisition period.

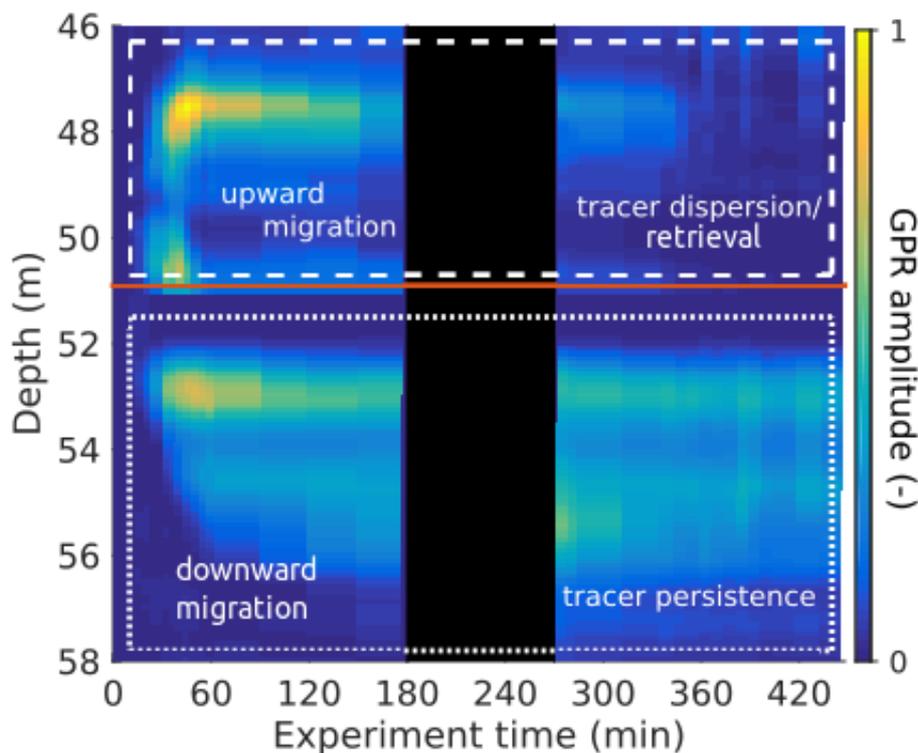

**Figure 8:** Plot of the GPR breakthrough curves computed for experiment III. The injection depth at 50.9 m is shown with the solid (red) line. The dashed and dotted rectangles correspond to the regions shown in Figure 7a. The GPR antenna batteries were charged between $t = 175$ and $t = 270$ min.

## 5. Interpretation

In this section, we discuss how the coupling of push-pull tracer tests and GPR imaging provides a means of estimating transport length scales, transport velocities, ambient flow and density effects.

*5.1. Transport length scales*

One major advantage of coupling push-pull experiments with GPR difference imaging is that the representative length scale of the experiment can be quantified, that is, how far did the saline tracer reach into the formation. For example, during the pushing and chasing phase of experiment II (Figure 5) with injection at 78.7 m, the saline tracer migrated approximately 8 m upwards from the injection location in a clearly defined trajectory. This is in contrast to the same phases of experiment III (Figure 7) with injection at 50.9 m, in which the saline tracer was seen to migrate less than 4 m from the injection location, but in several separate trajectories. Furthermore, different transport length scales are also highlighted within a single experiment when some of the tracer remains relatively close to the injection, while the rest of the tracer migrates further away. This information is crucial to assess the volume investigated by the tracer tests, as well as the characteristic fracture surface that is seen by the tracers, especially when considering reactive transport. It also helps to define the appropriate conceptual model for fracture flow and transport.

*5.2. Transport velocities and associated scale effects*

Another important advantage of single-hole GPR difference imaging is that inference about the tracer mean velocity can be made, since the tracer is mapped in both space and time. To demonstrate this, we consider the GPR breakthrough curves during the pushing and chasing period of experiments II, IV and V. These experiments were conducted at the same location and apart from minor changes in the experimental parameters (injection rate, tracer concentration and volume) they can be considered otherwise identical. We used the corresponding GPR breakthrough curves at each depth interval to compute a mean arrival time of the tracer that we defined as the time at which each curve crosses a noise threshold , set as 30% of the maximum value seen in the difference images. In Figure 9, we plot the travel distance as a function of the mean arrival time of the tracer along its complete trajectory after chasing (see Figure 5h for the complete trajectory), hence, the slope of this plot can be thought of as the mean velocity of the tracer. The overall agreement of the mean arrival times for the different experiments not only indicates that the push-pull experiments and the GPR monitoring results were repeatable, but also that calculating GPR breakthrough curves is a sound approach.

From this plot we can identify two flow regimes; for the first 3 meters the tracer velocity appears to be constant (at 0.18 m min$^{-1}$, expected for channelized flow) and for the next 4 meters the velocity is decreasing with a rate as expected for radial flow along a plane. For comparison, we plot the analytical solutions of advective transport within a 1D pipe and within a 2D fracture plane. We assume that the tracer reaches a distance of $x = 8$ m within $t = 45$ min and compute the curves $x = (8/45)\,t$ for the 1D solution and $x = (8/45^{1/2})\,t^{1/2}$ for the 2D solution.

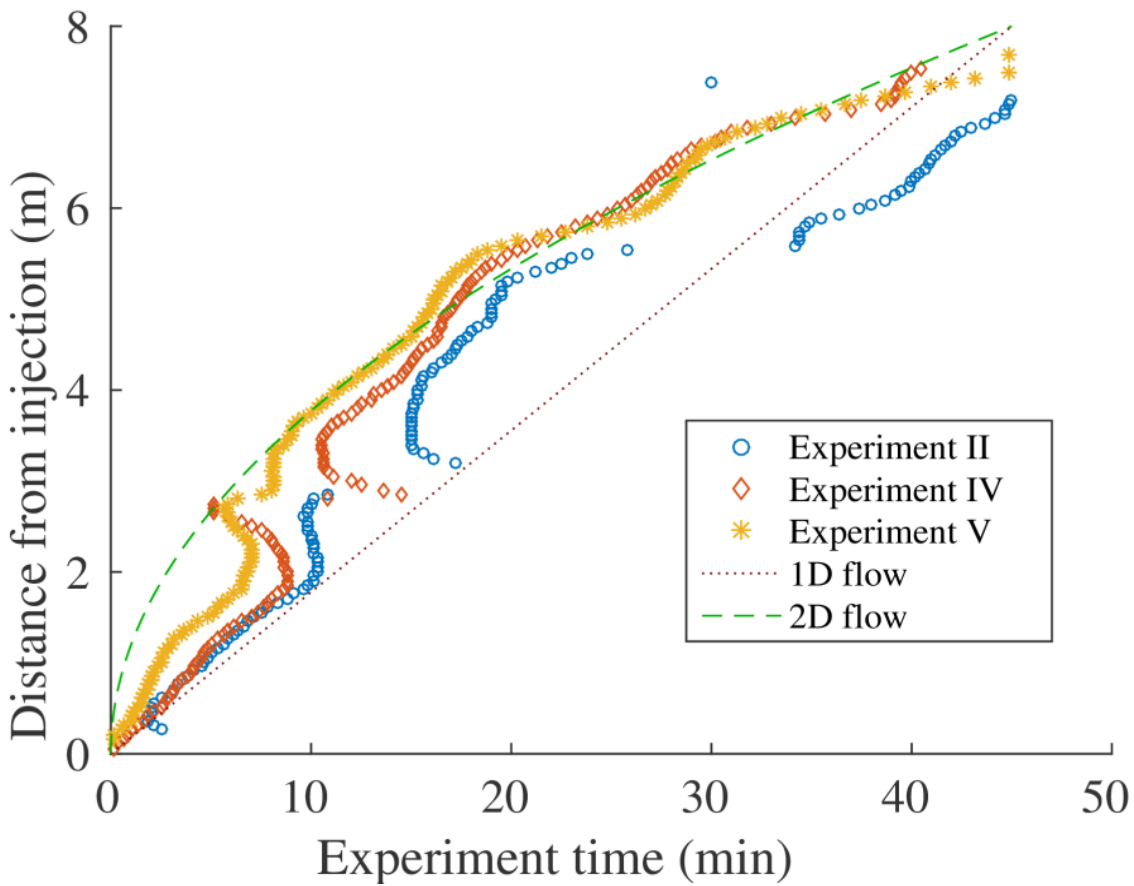

**Figure 9:** Travel distance versus time of the saline tracer for experiments II, IV and V, computed using the GPR breakthrough curves. The analytic solutions for 1D (channelized) and 2D (planar) flow are also shown, assuming the saline tracer travels a total distance of 8 m in 45 min.

*5.3. Geometric properties of transport flowpaths*

Another type of information that is lacking in classical push-pull experiments is information about the complexity of the pathways that the tracer follows, such as the number of fractures involved as well as their orientations and sizes. For example, borehole optical logs of B1 [*Belghoul*, 2007] along with prior push-pull and flowmeter tests suggest the presence of an open fracture with a dip angle of 15° at 78.7 m and another with a dip angle of 37° at 50.9 m [*Le Borgne et al.,* 2007]. From the difference sections obtained in experiment II (Figure 5), we see changes that originate from the injection at 78.7 m and extend vertically upwards at an angle steeper than 70°. This suggests that the sub-horizontal fracture that intersects the borehole is connected to a major sub-vertical fracture that is dominating the flow behavior during the experiment. Similarly, the difference sections for experiment III (Figure 7) suggests that the flow originating from the injection location at 50.9 m is dominated by two sub-vertical fractures on either side of the injection location. Since these fractures do not intersect the injection interval, a sub-horizontal fracture (not clearly visible in the difference sections) may have provided the necessary pathway for the tracer. The presence of a sub-horizontal fracture in this injection interval is in agreement with *Dorn et al.* [2011, 2013, 2012b].

*5.4. Observation of density effects and ambient flow*

It is important to assess the effect of density on the tracer distribution as we use a tracer with a high salinity contrast. In experiment I (Figure 4) we measured a decrease in the GPR breakthrough curve when the pumping rate is lowered (by a factor of 10) to inject the second tracer, and the curve only continued to increase when the previous pumping rate was resumed. Since the antennas were located 3.6 m above the injection interval, this decrease may have occurred because the lower pumping rate was not sufficient to push the tracer upwards, but instead a downward (density-driven) migration of the tracer occurred. Immediately after the previous pumping rate resumed, the tracer continued its upward migration and the GPR breakthrough curve peaks during the chasing period. A smaller peak is seen when the flow is reversed that results from the downward migration of the tracer towards the injection interval.

In experiment II (Figure 5) we saw an upward migration of the tracer during the pushing and chasing, followed by a sharp decrease in GPR difference amplitudes in the upper region (71 to 75 m) as soon as pulling begun (compare Figs. 5h and 5i). As already suggested in the previous paragraph, the sharp decrease partly occurs due to the fast downward migration of the tracer. However, some of the visible tracer may have been dispersed into sub-horizontal fractures to which our GPR data are largely insensitive. Ambient flow in these fractures may have led to the low mass recovery for this experiment (Table 1).

In experiment III (Figs. 7 and 8) we see different behavior of the tracer in the upper (45 to 51 m depth) and lower (51 to 59 m depth) regions relative to the injection interval. While the upper region shows strong amplitude changes early on, the amplitudes are quickly diminished during the resting time, possibly due to ambient flow in this location that causes the tracer to be quickly dispersed. As soon as pulling started, the GPR breakthrough curves from this region are quickly attenuated to the background level. On the contrary, the lower region exhibits persistent amplitude differences with respect to the reference throughout the whole experiment, indicating that this fracture is perhaps not strongly affected by ambient flow. The same analysis performed on experiment IV showed that the tracer persists during the resting time in the bottom part of the fracture (from the injection depth to 76 m depth in Figure 5) while in the top part (from 76 m to 71 m in Figure 5) the tracer is quickly dispersed.

Another effect that is visible in both the difference sections (Figure 7) and in the GPR breakthrough curves (Figure 8) is the vertical spreading of the saline tracer, mainly during the resting time, in the fracture found in the lower region. This is most likely due to density effects since the saline tracers used have strong density contrasts to the surrounding fluid. Note that saline tracers are needed to image tracer transport with GPR difference methods but the salinity could be reduced to minimize density effects, while still conserving sufficient contrasts in the GPR images.

*5.5. Physical meaning of early tracer breakthrough*

To investigate the effect of density on the push-pull results we simulated experiment II, using the measured flow rates and converted the electrical conductivity into salinity [*Sen and Goode*, 1992] using the reference temperature of 25ºC used by the CTD diver. For the simulation, we modified a finite volume solver for flow-and-transport in 2D porous media [*Künze and Lunati*, 2012]. The solver allows for density effects by incorporating the force of gravity into the solution of the flow velocity. We adjusted the permeability to account for the transmissivity of a parallel plate fracture through the cubic law [*e.g., Lunati et al.,* 2003] and scaled the porosity by the aperture to account for the volume occupied by the tracer. To allow for inclinations of the fracture we multiply the gravity term by a component that is equal to the sine of the dip angle. Flow in a fracture with zero dip angle is therefore not affected by gravity. Guided by the changes shown in the GPR sections of experiment II (Figure 5), we defined as a modeling domain a fracture with no-flow boundaries on the bottom and sides and with an open boundary on the top so that tracer movement would be primarily upwards. To avoid any tracer reaching the open boundary we kept the total volume of the fracture two times larger than the volume of injected fluid in the domain. We then performed a local optimization search (using the golden section search algorithm) to match the peak arrival in the data with a uniform aperture. Note that this simplified model that ignores dispersion, fracture heterogeneity, and ambient flow only serves to highlight the impact of density and fracture dip on the peak arrival time, and that no emphasis has been placed on fitting the tail of the tracer breakthrough curve and honoring the recovered mass.

Using this simplified model we attempted to reproduce the early arrival of the tracer seen in the tracer breakthrough curve (Figure 6). We allowed the fracture dip to vary between 0° (no density effect) and 90° (maximum density effect) in increments of 10°. The width of the fracture, for a given aperture, was chosen such that the volume of tracer at the end of the chasing phase would occupy a length of 8 m along the fracture. This constraint was motivated by the GPR difference sections (Figure 5h). In Figure 10a, we plot the optimal aperture for each dip angle along with the misfit in peak arrival time between the simulations and the experiment. It is evident that high dip angles are required to fit the early arrival times obtained in the experiment, with an optimal effective aperture of roughly 5 mm. The simulations not only indicate that the effect of density should be modeled when using highly saline tracers, but more importantly that the constraints on the tracer trajectory obtained from the GPR difference images are necessary to obtain realistic estimates. We also plot (Figure 10b) the measured and (optimal) simulated curves for fractures with 0 and 70 degrees dip angle. Note that we use a normalized concentration for comparison purposes, since our modeling does not account for processes that would reproduce the lower mass recovery seen in the field data.

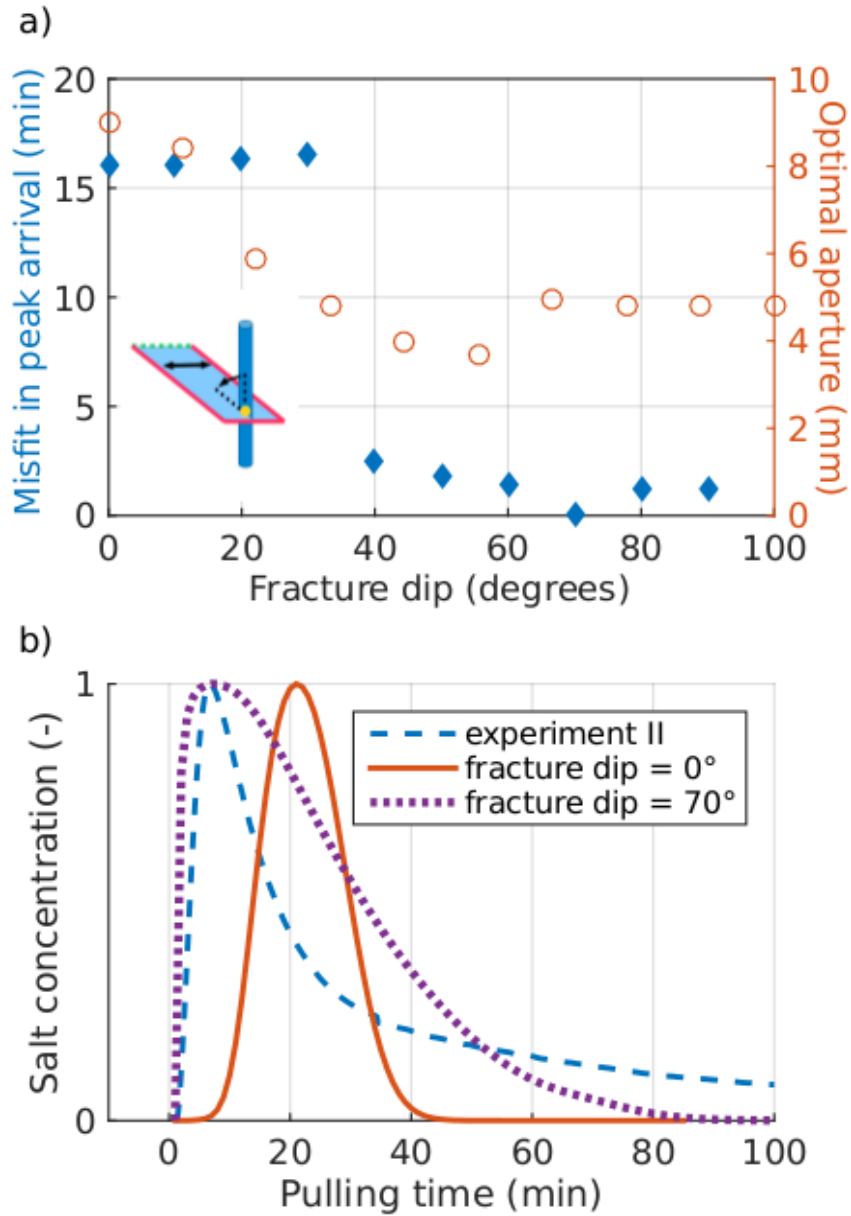

**Figure 10:** (a) Plot of the misfit (left axis) between the simulated and measured peak arrival time of the saline tracer in experiment II (see subfigure for the modeling setup). Injection was performed at the bottom of a uniform fracture (yellow circle) with no-flow boundaries (red solid lines) on the bottom and two sides. The distance between the two sides (double arrow) was adjusted so that the tracer would reach a length of 8 m at the end of the chasing period while the open boundary at the top (dashed green line) was kept far from the tracer. The simulations were performed for 10 dip angles (inclined arrow), between 0° and 90°, and each simulation was optimized to obtain the best fitting mean aperture (right axis). The (normalized) best fitting simulation for 0° and 70°, along with the measured concentration for experiment II are shown in (b).

*5.6. Breakthrough curve tailing*

As discussed in section 2, previous fluorescent dye experiments measured power law breakthrough curve tailing, such as $c(t) \sim t^{-1-b}$ at late time, with *b* ranging from 0.75 to 1, and concluded to a limited role of matrix diffusion in this low porosity granite [*Kang et al.,* 2015]. The tailing behavior measured here is more pronounced, showing a behavior close to $c(t) \sim t^{-1}$. This strong tailing is thus likely induced by density effects affecting the transport of the saline tracer used in these experiments.

Our results (Figs 4, 6 and 8) suggest a strong link between the tracer and GPR breakthrough curves. In Figure 11 we plot the tracer breakthrough curves for experiments I, II and III. The repeatability of the push-pull experiments is evident in the tracer breakthrough curves for experiments I and II (same experimental parameters), while the tracer breakthrough curve for experiment III (a different injection location) indicates an early arrival time and a longer tail compared to the other two experiments. During the later part of the pulling period (between $t = 100$ and $t = 200$ min) the concentration of recovered tracer is higher in experiment III than in the other two experiments. This agrees well with the GPR difference sections (Figure 7) and the GPR breakthrough curves (Figure 8), which show persistent reflections arising from the lower region. The tailing in the tracer breakthrough curve, in light of the GPR results, can thus be attributed to tracer that has migrated downwards during the resting time.

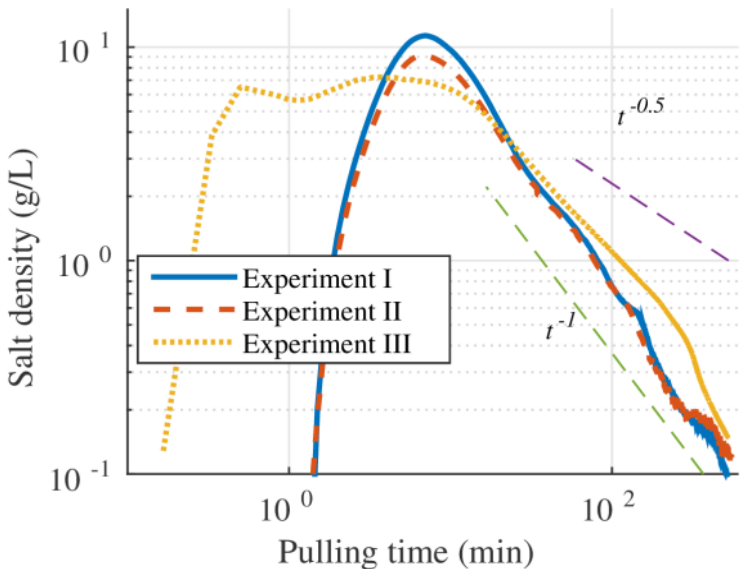

**Figure 11:** Measured tracer breakthrough curves for experiments I, II and III presented in Table 1 along with plots for arrival times with exponents -1 and -0.5.

**6. Conclusions and Outlook**

Combined push-pull tracer tests and single-hole GPR monitoring experiments with a saline tracer provides important insights into a range of transport phenomena that can be observed in situ in fractured rocks. This includes characteristic length scales of transport, scale effects in transport velocities, ambient flow and density driven migration, and trapping effects. The information contained in the single-hole GPR reflections helps to gain physical understanding of the processes that are responsible for the observed tracer breakthrough curves.

During a fixed antenna experiment we measured a GPR breakthrough curve that was consistent with the tracer breakthrough curve and provided a means to investigate the rise and decay of tracer concentrations at different distances along the main fracture flow paths. This relationship was used to deduce a mean velocity of the saline tracer during the moving antenna experiments. During the push phase we observed a scale effect in tracer displacement, with a transition from linear to radial flows. This suggests that a significant flow channeling controls transport until a characteristic scale of about 3 meters, above which it follows a radial behavior. Our results demonstrate that GPR monitoring combined with push-pull saline tracer tests is useful to image flow channeling in fractured rocks, at least in the near field around boreholes.

We used a 2D flow and transport solver, modified to simulate flow in a parallel plate fracture, to show that, when density effects are significant, the peak tracer breakthrough time is highly sensitive to the fracture dip. Our modeling results suggest that a large dip is necessary to reproduce the early tracer arrival measured in the field and this is in agreement with the large dip seen in the GPR difference sections.

These experimental results offer new perspectives for combined hydrogeophysical modeling of fractured rock formations. In particular, the new insights obtained may be very useful for interpreting reactive tracer tests, often performed in push-pull configuration [*Istok*, 2012] since reaction rates are expected to depend on the spatial tracer distribution and its surface of exposure to the rock. To obtain quantitative information on tracer mass distribution in situ, we plan to couple flow-and-transport modeling with a recently developed approach to simulate GPR reflections from saline-occupied fractures of arbitrary orientation and aperture [*Shakas and Linde*, 2015].

**Acknowledgements**

This research was supported by the Swiss National Science Foundation under grant 200021-146602 and by the French National Observatory H+ (hplus.ore.fr/en). We would like to thank Pavel Tomin and Ivan Lunati for their help with the MaFloT simulations, Antoine Armandine les Landes for fruitful discussions on modeling density effects and Valentin Guilbaut for his indispensable help in the field. The comments provided by Associate Editor Lee Slater, George Tsoflias and two anonymous reviewers helped to significantly improve the manuscript. The data and input files required to reproduce the results are available from the first author upon request.